# Title: Survival Function Analysis of Planet Size Distribution with GAIA Data Release 2 Updates

**Authors:** Li Zeng*[1], Stein B. Jacobsen[1], Dimitar D. Sasselov[2], Andrew Vanderburg[3]

**Affiliations:**

[1]Department of Earth & Planetary Sciences, Harvard University, 20 Oxford Street, Cambridge, MA 02138.

[2]Harvard-Smithsonian Center for Astrophysics, 60 Garden Street, Cambridge, MA 02138,

[3]Sagan Fellow, University of Texas at Austin

Correspondence to: astrozeng@gmail.com

**Abstract:**

Applying the survival function analysis to the planet radius distribution of the *Kepler* confirmed/candidate planets, we have identified two natural divisions of planet radius at 4 Earth radii ($R_\oplus$) and 10 $R_\oplus$. These divisions place constraints on planet formation and interior structure model. The division at 4 $R_\oplus$ separates small exoplanets from large exoplanets above. When combined with the recently-discovered radius gap at 2 $R_\oplus$, it supports the treatment of planets in between 2-4 $R_\oplus$ as a separate group, likely water worlds. For planets around solar-type FGK main-sequence stars, we argue that 2 $R_\oplus$ is the separation between water-poor and water-rich planets, and 4 $R_\oplus$ is the separation between gas-poor and gas-rich planets. We confirm that the slope of survival function in between 4 and 10 $R_\oplus$ to be shallower compared to either ends, indicating a relative paucity of planets in between 4-10 $R_\oplus$, namely, the sub-Saturnian desert there. We name them transitional planets, as they form a bridge between the gas-poor small planets and gas giants. Accordingly, we propose the following classification scheme: (<2 $R_\oplus$) rocky planets, (2-4 $R_\oplus$) water worlds, (4-10 $R_\oplus$) transitional planets, and (>10 $R_\oplus$) gas giants.

**Method:**

### (1) Survival Function Analysis

The survival function (Clauset, Rohilla Shalizi, & J Newman, 2009; Feigelson & Nelson, 1985; Virkar & Clauset, 2014), also known as the complimentary cumulative distribution function (cCDF), is defined in this context as the number of planets above a given radius, versus radius in a log-log plot, of the *Kepler* confirmed/candidate planets: **Figure 1** shows 4433 of them from Q1-Q17 DR 25 of *NASA* Exoplanet Archive (Akeson et al., 2013; Thompson et al., 2017), and a subset of 1861 of them from the California-Kepler Survey (CKS) with improved host-stellar parameters (Fulton et al., 2017), and 4268 with improved planet radii from GAIA Data Release 2 (DR2) (Berger, Huber, Gaidos, & van Saders, 2018; Gaia Collaboration et al., 2018; Lindegren et al., 2018).

Definition of Survival Function:
SF (Survival Function) = 1- CDF (Cumulative Distribution Function) = 1 - Integral of PDF (Probability Density Function)
Differentiate SF, one gets the PDF (Probability Density Function).

The survival function (SF) can tell apart different distributions. Comparing to the probability density function (PDF), it has the advantage of overcoming the large fluctuations that occur in the tail of a distribution due to finite sample sizes (Clauset et al., 2009). For example, on a log-log plot of survival function, power-law distribution appears as a straight line, while normal, log-normal, or exponential distributions all have a sharp cut-off (upper bound) in radius. This plot is also known as the rank-frequency plot (Newman, 2005). This approach identifies the boundaries separating different regimes of distributions in the data.

As shown in **Figure 1**, the breaks (**3.9 ± 0.1 R⊕** and **10.3 ± 0.1 R⊕** to be exact) in the survival function are the natural (model-free) boundaries of different regimes of planets, in addition to the gap in exoplanet radius distribution detected around **2 R⊕** (Berger et al., 2018; Fulton et al., 2017; Fulton & Petigura, 2018; Van Eylen et al., 2017; Zeng, Jacobsen, Hyung, et al., 2017; Zeng, Jacobsen, & Sasselov, 2017). Thus, we propose the following classification schemes based on planet radius:

- **<4 R⊕**, small planets. They can be further divided into two sub-groups: **<2 R⊕** and **2-4 R⊕**. The small planets are generally gas-poor, with gaseous envelope mass fraction ($f_{env}$ = $M_{env}/M_{planet}$) less than about 5~10%. The upper bound of whether ~5% or ~10% depends on the assumptions of core mass, core composition (core here refers to the solid part of the planet), envelope thermal profile and envelope metallicity. The details of calculations can be seen in (Ginzburg, Schlichting, & Sari, 2016, 2017; Lopez & Fortney, 2014). If one assumes a water-rich core like that of Uranus or Neptune, then this upper bound is more like ~5%. If one assumes a rocky core, then this upper bound is more like ~10%.
- **4-10 R⊕**, transitional planets. Statistics of the *Kepler* confirmed/candidate planets shows that this group of planets follows a power-law distribution as: $dN \propto R^{-\alpha} * dR$, where $\alpha \approx 2$ (1.9±0.1 to be exact). The power index $\alpha$ in this radius range is shallower than ranges above and below, which means a relative paucity of planets per logarithmic interval of radius. This confirms the <u>sub-Saturnian desert</u>. We name them "transitional planets" as they form a bridge between small exoplanets (<4 R⊕) and gas giants (>10 R⊕).
- **>10 R⊕**, gas giants. They are dominated by $H_2$-He in their bulk composition and are massive. They include Jupiter-sized planets, brown dwarfs, and even small stars.

**(2) Error Analysis**

**Figure 2** applies a Monte-Carlo method to determine the uncertainty in the survival function. For example, each of the planet radii has some best-fit value with some uncertainty (for example, 1 ± 0.1 R⊕, 2.3 ± 0.05 R⊕, etc). **Figure 1** so far calculates the survival function with the best-fit values (in those two examples, we would use 1, and 2.3, for example).

**Figure 2** then randomly draws a number for each planet from the asymmetric Gaussian distribution centered at the mean value with a width equal to the uncertainty in the plus or minus direction in the radius measurement. For those two examples, we might randomly draw 1.03 and 2.34 $R_\oplus$ and calculate the survival function with these newly drawn radius measurements. Then, repeat that whole process 100 times, and calculate 100 survival functions. This gives a sense as to the uncertainty in the survival function itself.

In **Figure 2**, various cuts are performed on the datasets (both KOI and CKS), with the same selection steps and criteria presented in (Fulton et al., 2017). The sequential cuts come at the expense of losing some potentially valuable data points and suffer more and more small number statistics and fluctuations towards larger radius. So, there is a trade-off. As shown in **Figure 2**, the breaks at 4 $R_\oplus$ and 10 $R_\oplus$ are robust. And the identification of the slope of about -1 in the survival function of planets in between 4-10 $R_\oplus$ is also robust.

In more detail, beyond excluding the false positives, all the subsequent cuts are throwing away mostly valid planets and may have the risk of introducing artificial features into the sample. For example, the cut at 100-day orbital period may alter the overall number ratio of small versus large exoplanets. As demonstrated by (Zeng, Jacobsen, Sasselov, & Vanderburg, 2018), large exoplanets (>4 $R_\oplus$) are relatively depleted inside 0.4 AU compared to small exoplanets (<4 $R_\oplus$). More specifically, large exoplanets (>4 $R_\oplus$) have a different statistical distribution in their semi-major axis or period, which is uniform in the square-root of semi-major axis, compared to small exoplanets (<4 $R_\oplus$), which is uniform in the logarithm of semi-major axis, within 0.4 AU or 100-day orbit, up to the inner threshold of 0.05 AU. Therefore, making a cut at 100-day orbital period will lose quite a few of large exoplanets (>4 $R_\oplus$).

Anyway, if one applies the strictest criteria and adopts all the cuts, as shown in **Figure 2**, the general trend of survival function is clear with the two breaks in the power-law, and the identification of the slope in between 4 and 10 $R_\oplus$ to be shallower than either end, indicating a relative paucity of planets in between 4-10 $R_\oplus$, namely, the sub-Saturnian desert there.

**(3) Analysis of updated planets' radii and insolation from GAIA DR2**

**Figure 3** shows the 2-dimensional scatterplot of planet radius-versus-insolation from GAIA DR2 updates (Berger et al., 2018), and a smooth kernel histogram derived from which using the Sheather-Jones bandwidth selector. The contours of the smooth kernel histogram in **Figure 3** confirm the boundaries at 2, 4 and 10 $R_\oplus$.

The radius gap at $2=10^{0.3}$ $R_\oplus$ can be viewed as the valley separating the two planet populations below and above. This gap runs perpendicular to the radius-axis. Superficially, this gap disappears around ~100 Earth fluxes ($F_\oplus$), likely an artifact due to geometric transit probability. It is not because the gap does not exist at lower flux or larger orbital distances, but because the geometric transit probability decreases as $(1/\mathbf{a})$ where $\mathbf{a}$ is the orbit semi-major axis, so it looks as if the population below the gap diminishes and the gap vanishes. This point will be demonstrated in forthcoming analysis that once the geometric transit probability is corrected, both populations (1-2 $R_\oplus$ versus 2-4 $R_\oplus$) are nearly flat (log-uniform) from the inner edges (the attenuation of both populations towards higher fluxes) out to at least 1 AU or 1 $F_\oplus$. Also, the gap is partially filled-in (Berger et al., 2018; Fulton & Petigura, 2018), contradictory to earlier speculations of an observationally under-resolved and forbidden region of planet sizes (Fulton et al., 2017; Van Eylen et al., 2017).

The break at $4=10^{0.6}$ $R_\oplus$ can be viewed as a rapid decrease in the probability density of the population above the gap towards larger radii. Therefore, 4 $R_\oplus$ is the lower-bound of the sub-Saturnian desert. The break at $10=10^{1.0}$ $R_\oplus$ is the upper-bound of the sub-Saturnian desert.

### (4) Completeness corrections

So far, the SF (**Figure 1** and **Figure 2**) and Histogram (**Figure 3**) do not represent the true distribution of planets, until we make completeness corrections for and **(1) pipeline incompleteness** and **(2) geometric transit probability**.

### (4.1) pipeline incompleteness

The pipeline incompleteness corresponds to a threshold of detectability in signal-to-noise ratio from a duration-limited transit survey. This threshold can be translated into a threshold radius for a given orbital period. This threshold radius varies roughly like orbital period $\mathbf{P}^{(1/6)}$ out to twice the duration of the transit survey (Pepper, Gould, & Depoy, 2002; Winn, 2018).

As pointed out by (Fulton et al., 2017), the pipeline completeness corrections are generally small. The geometric transit probability term dominates the corrections in most parameter space explored. When they divide planets into two bins of 1.0-1.75 $R_\oplus$ and 1.75-3.5 $R_\oplus$, they found that the mean pipeline completeness for each case is 86% and 96%. Therefore, in our manuscript, we choose to focus on correcting the geometric transit probability.

This threshold radius is less than 4 $R_\oplus$ for planets with orbital period less than 300 days. Therefore, it should not affect the identification of breaks in the power-law at 4 and 10 $R_\oplus$ in the survival function analysis.

### (4.2) geometric transit probability

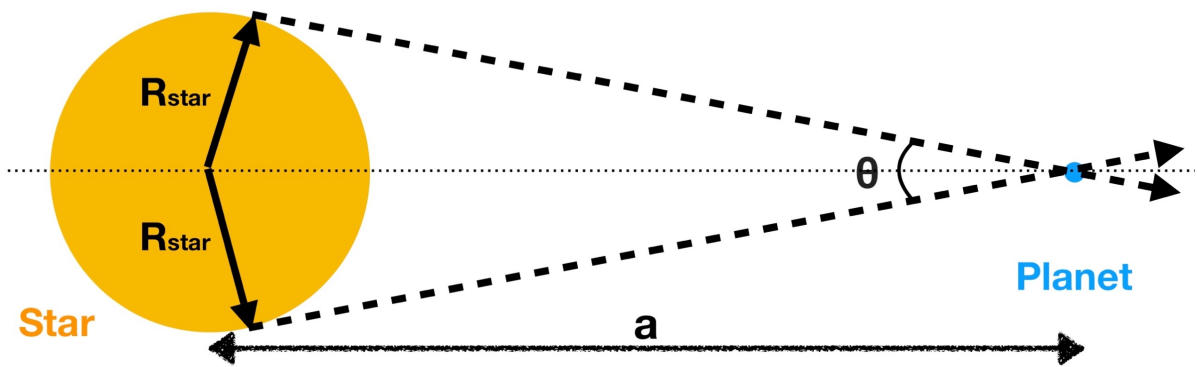

**Diagram 1**. Schematic diagram showing the geometry of transit, assuming the planet is small compared to the size of the star.

The **geometric transit probability** corresponds to a planet orbiting further from its host star has less chance of being aligned with our line-of-sight and thus less chance of transiting, see illustration above. If we assume the orientation of orbital plane in space is random, then this **probability of transit** $= \frac{(R_{star} + R_{planet})}{a \cdot (1 - e^2)}$ , (Winn, 2010). Since $R_{planet} \ll R_{star}$, and $R_{star} \approx R_{sun}$ (most host stars in the *Kepler* catalog are solar-type FGK main-sequence stars), and assume eccentricity *e* is small, then **probability of transit** $\approx \frac{R_{sun}}{a}$. To correct for this **probability of transit** to return the true distribution, the PDF needs to be multiplied by a factor of $\frac{a}{R_{sun}}$.

This geometric correction can be applied to **Figure 3** since it contains the flux and hence the orbital distance information simultaneously with the radius information. Both pieces of information are needed together to make the corrections.

But first, let's take a look at two vertical slivers (1-2 $R_\oplus$ versus 2-4 $R_\oplus$) enclosing the majorities of the two planet populations above and below gap in **Figure 3**, and project them to the vertical flux-axis as one-dimensional histograms shown in **Figure 4**. This projection demonstrates that the two planet populations overlap significantly in the flux-dimension. They show similar trend of attenuation towards higher flux, except shifted in flux by a factor of 3~4.

To strengthen this observation, we choose narrower ranges of planet radius selecting the peaks of the two populations and restrain our sample to only main-sequence host stars within the temperature range of 5000-6500 Kelvin (see **Figure 5**). This makes geometric-transit-probability correction easier. The smooth curves in **Figure 5** are smooth kernel fit to the underlying histogram. They can now be corrected for the geometric transit probability, in order to reproduce the approximate true flux-distribution of planets in each radius bin. We know that flux f:

$$f = \frac{L_*}{4\pi a^2} = \frac{4\pi R_*^2 \sigma T_*^4}{4\pi a^2} = \left(\frac{R_*}{a}\right)^2 \sigma T_*^4$$

So,

$$\left(\frac{f}{f_\oplus}\right) = \left(\frac{R_*/R_\odot}{a/AU}\right)^2 \left(\frac{T_*}{T_\odot}\right)^4 = \left(215 \cdot \frac{R_*}{a}\right)^2 \left(\frac{T_*}{T_\odot}\right)^4$$

Thus,

$$Transit\ Probability \approx \left(\frac{R_* + R_p}{a}\right) \approx \frac{R_*}{a} \approx \frac{1}{215} \cdot \left(\frac{f}{f_\oplus}\right)^{1/2} \cdot \left(\frac{T_*}{T_\odot}\right)^{-2}$$

Since we consider only the temperature range of 5000-6500 K, which deviates from $T_\odot = 5777$ K by at most $\pm 13\%$ in both directions, and we plot in the logarithmic scale of flux, the temperature-depended term can be dropped for now, for convenience of calculation.

Therefore, we divide the PDF in **Figure 5** by this transit probability, to come up with a transit-probability-corrected PDF to approximate the true flux-distribution of planets for each planet population, shown in **Figure 6**.

**Discussion:**

The two planet populations above and below the radius gap at ~2 $R_\oplus$ overlap significantly in the flux-dimension (by at least two-orders-of-magnitude) (**Figure 4**, **Figure 5**, and **Figure 6**). If planets of (1-2 $R_\oplus$) mainly result from evaporation of planet of (2-4 $R_\oplus$), then the low in PDF of one population should be the high in PDF of the other. This is not seen. Furthermore, the continuation of (1-2 $R_\oplus$)-planet population into the low-flux/long-orbital-period region (≲100 $F_\oplus$) suggests that at least some of them are intrinsically there by (formation+migration) but not by evaporation.

The exact match of the shape (functional form) of the attenuation in PDF towards higher-flux for both planet populations, below and above the gap, indicates that this attenuation is likely caused by the same physical mechanism. This physical mechanism is effective in attenuating the PDF of planets of (1.3-1.8 $R_\oplus$), many of which are consistent with pure-rocky composition without significant envelope according to the limited mass measurements available from the radial velocity follow-ups (Zeng, Jacobsen, Sasselov, Vanderburg, et al., 2018).

To summarize:
The PDF in log-flux of (1.3-1.8 $R_\oplus$)-planet-population attenuation starts at ~100 $F_\oplus$, reduces to half the amplitude at ~400 $F_\oplus$, and cuts off completely at 5000 $F_\oplus$.
The PDF in log-flux of (2-3 $R_\oplus$)-planet-population attenuation starts at ~25 $F_\oplus$, reduces to half the amplitude at ~100 $F_\oplus$, and cuts off completely at 2000 $F_\oplus$.

For comparison, the prediction from the gas dwarf evaporation hypothesis (see **Figure 7**) shows almost no planets of (2-4 $R_\oplus$) to reside inside 10-day orbit (≳120 $F_\oplus$). That is simply because if planets of (2-4 $R_\oplus$) are gas dwarfs: a rocky core surrounded by a few mass percent $H_2$-He-dominated gaseous envelope, then they cannot withstand that high-level of flux. Contrarily, **Figure 3-6** show that there are plenty of planets of (2-4 $R_\oplus$) inside 10-day orbit (≳120 $F_\oplus$). Some of them can withstand up to hundreds and even ~1000 $F_\oplus$. If they are water worlds and have $H_2O$-dominated envelope, then they may withstand that much flux and reach certain equilibrium state.

**Figure 7** also shows a quickly diminishing population of planets of (1.3-1.8 $R_\oplus$) beyond ~20-day orbit (≲ 50 $F_\oplus$). This reveals a general caveat of the gas-dwarf-evaporation-by-host-star models that they fail to reproduce the long-period rocky planet population. The ratio of PDFs between the two populations, below and above the gap, as predicted by these models, for example PDF[(1.3-1.8 $R_\oplus$)]/PDF[(2-3 $R_\oplus$)], is strongly dependent upon orbital period or flux or semi-major axis. However, **Figure 6** shows that both populations are nearly log-uniform (flat) in the low flux region (≲100 $F_\oplus$). So their PDF ratio stays nearly constant for at least two-orders-of-magnitude in flux. This is a strong indication that both planet populations, below and above the radius gap ~2 $R_\oplus$, exist intrinsically, but not due to the influence of their host stars.

**(1) Speculation on the origin of the kink at 4 $R_\oplus$**
It is interesting, perhaps not by coincidence, that our own solar system happens to have planets near these two kinks (~4 $R_\oplus$ for Uranus and Neptune, ~10 $R_\oplus$ for Jupiter and Saturn) of the SF. It is generally known that Uranus and Neptune are icy giants, that is, their interior compositions are dominated by a mixture of ices, surrounded by about a few up to 10 percent $H_2$/He envelope by mass (Hubbard et al., 1991; Podolak, Podolak, & Marley, 2000).

(Ginzburg et al., 2016) suggests that the intrinsic luminosity coming out of the cooling cores of young planets themselves can gradually blow off small envelope ($f_{env} \lesssim 5\%$) without the need of stellar irradiation. And planets of $f_{env} \gtrsim 5\%$ can largely retain their envelopes over billion-year timescale. This is one possibility to explain the origin of the boundary at **4 $R_\oplus$**. (Ginzburg et al., 2017) applies this argument to explain the radius gap at 2 $R_\oplus$ by convolving this bi-modal envelope mass fraction with a core mass distribution centered around 3-5 Earth masses ($M_\oplus$) (similar to what was assumed in (Owen & Wu, 2017)). Instead, this argument works better to explain the boundary at **4 $R_\oplus$**, if the same envelope mass fraction is convolved with slightly larger core sizes of 2-2.5 $R_\oplus$, consistent with water-rich cores, for core masses of 5-20 $M_\oplus$, consistent with current mass measurements from the radial-velocity (Zeng, Jacobsen, Sasselov, Vanderburg, et al., 2018). If this is true, then **4 $R_\oplus$** separates core-dominated planets from gaseous-envelope-dominated planets. The sizes of water-rich cores of 5-20 $M_\oplus$ are 2-2.5 $R_\oplus$, so **4 $R_\oplus$** is the divide of two regimes: envelope thickness $\lesssim$ core radius versus envelope thickness $\gtrsim$ core radius (Ginzburg et al., 2017).

Earth once had a $10^3$-$10^4$ bar primordial $H_2$/He envelope when the solar nebula gas disk was around, preserved as noble gas signature in present-day Earth mantle due to early exchange/in-gassing of this primordial gaseous envelope with the early Earth magma ocean (Harper & Jacobsen, 1996). This primordial envelope is still small in its mass fraction compared to the $f_{env} \sim 5\%$ threshold (Ginzburg et al., 2016). Thus, Earth was not able to hold on to this envelope after disk dispersal, due to a combination of spontaneous-driven and stellar-driven losses. Similar scenario can apply to many other cores, if they do not grow fast enough to reach the critical mass to accrete gas efficiently, then after the gas disk dispersal, or after fast inward migration, some of them would lose their primordial envelope eventually.

## (2) Comparison to previous works
### (2.1) Host-star metallicity

The divide at **4 $R_\oplus$** confirms the work by (Buchhave et al., 2014) by looking at the host stars' metallicities. They show that the host stars of planets with radius larger than 4 $R_\oplus$ are metal-enriched ($0.18 \pm 0.02$ dex), compared to the host stars of planets with radius smaller than 4 $R_\oplus$ which generally have solar-like metallicities. (Winn et al., 2017) studies the ultra-short-period (USP, P< 1 day, R<2 $R_\oplus$) planets and shows that the metallicity distributions of USP planets and hot-Jupiter hosts are very different, suggesting that USP are not dominated by the evaporated cores of hot Jupiters. The metallicity distribution of stars with USP is indistinguishable from that of stars with short period (1-10 day) planets of (2-4 $R_\oplus$). From the metallicity correlation, we infer that **4 $R_\oplus$** is the divide of gas-poor and gas-rich planets. Planets above 4 $R_\oplus$ have substantial gaseous envelopes and their host stars are statistically metal-enriched compared to solar metallicity. Thus, substantial envelope is correlated with enhanced metallicity: this is the beginning point of the giant-planet-metallicity correlation (Fischer & Valenti, 2005; Wang & Fischer, 2015).

**(2.2) Comparison to previous claimed rapid drop-offs in planet fraction at lower radii**

Figure 7 of (Fressin et al., 2013) shows the radius domain of what they call "Small Neptunes" (2-4 $R_\oplus$), corresponding to the boundaries and identification of water worlds in this paper. On page 13, they point out that "the increase in planet occurrence towards smaller radii from these objects is very steep". In their analysis, they attempt to place a boundary within "Small Neptunes" (2-4 $R_\oplus$) at 2.8 $R_\oplus$. But this is done in an artificial way, quotation: "we find that dividing the small Neptunes into two subclasses (two radius bins of the same logarithmic size: 2-2.8 $R_\oplus$ and 2.8-4 $R_\oplus$), we are able to obtain a much closer match to the KOI population (K-S probability of 6%) with similar logarithmic distributions within each sub-bin as assumed before". The key here is that they have <u>assumed</u> logarithmic distribution of planet sizes within each planet category. But this assumption is not valid according to what data show in the SF analysis that the planet size distribution fits well with piece-wise power-law. SF analysis also shows no obvious change of slope at 2.8 $R_\oplus$. So, the boundary at 2.8 $R_\oplus$ is simply chosen because it sits at the logarithmic mid-point between 2 and 4 $R_\oplus$.

In (Petigura, Howard, & Marcy, 2013), they choose the same radius bins as that of (Fressin et al., 2013), see their Figure 3: a histogram showing the counts of planets within each bin. As explained above, 2.8 $R_\oplus$ is chosen as the logarithmic mid-point between 2 $R_\oplus$ and 4 $R_\oplus$. The fact that planets in between 2-2.8 $R_\oplus$ are many more than 2.8-4 $R_\oplus$ is simply a manifestation of the steep slope of the power-law distribution in this radius range, that is, the population quickly diminishes towards larger radii, instead of a change of slope at 2.8 $R_\oplus$. (Silburt, Gaidos, & Wu, 2015) also adopts the same binning as (Fressin et al., 2013) and (Petigura et al., 2013). In all these previous analyses, 2.8 $R_\oplus$ is chosen but not detected.

**(3) Slope of power-law distribution**

The probability distribution of the transitional planets (4-10 $R_\oplus$) is best fit to a power-law: $dN \propto R^{-\alpha} * dR$, where the power index $\alpha \approx 2$ (1.9±0.1 to be exact). It likely implies a power-law distribution in planet mass as well, but this of course depends on the exact mass-radius relationship of planets in this radius range. Many natural phenomena follow a power-law distribution with power index $\alpha$ typically in the range of 2~3 (Clauset et al., 2009; Newman, 2005). The examples include the frequency of use of words, magnitude of earthquakes, diameter of moon craters, population of US cities, etc. (Newman, 2005) There are many mechanisms proposed for generating power-law distribution in nature, such as via preferential attachment, multiplicative processes, random walks, phase transitions and critical phenomena, etc. (Mitzenmacher, 2003; Newman, 2005). In our case of transitional planets, the physical mechanism accounting for the power index of $\alpha \approx 2$ could be a combination of the equations of states (EOS) of $H_2/He$ envelope in the planet interiors and the formation and growth processes of these planets.

The SF of small exoplanets (<4 $R_\oplus$) suggests an overabundance of them over the extrapolation of the power-law SF of transitional planets (4-10 $R_\oplus$). The PDF of small exoplanets (<4 $R_\oplus$) in planet radius can be fit to two log-normal distributions with two peaks at ~1.5 $R_\oplus$ and ~2.5 $R_\oplus$ (Berger et al., 2018; Fulton et al., 2017; Fulton & Petigura, 2018; Van Eylen et al., 2017; Zeng, Jacobsen, Hyung, et al., 2017; Zeng, Jacobsen, Sasselov, Vanderburg, et al., 2018).

**(4) Rocky versus water-rich cores**

There is a clear bi-modal distribution of the densities of solar system satellites (**Figure 8**), indicative of two generic types of objects in our own solar system: rocky versus icy. The rocky ones are primarily rocky in their composition with a maximum of a few percent ices by mass, while the icy ones are composed typically of one-third to one-half of ices or even more, where the ices are dominated by $H_2O$-ice. So, the amount of ices on any object is not arbitrary.

One example is the Galilean moon system around Jupiter, where Io and Europa have densities >2.5 g/cc (even though Europa has a surface ice-ocean-layer, it is thin compared to its radius and comprises a small fraction of its bulk mass), while Ganymede and Callisto have densities <2 g/cc, consistent with a mass fraction of ice of 1/2. Saturn's largest moon Titan, Neptune's largest moon Triton, Pluto and Charon, are all estimated to be made of one-third to one-half water ices and the rest as rocky material based on their bulk densities. We argue that the bi-modal distribution that we see in exoplanets is simply a scaled-up version of this picture.

The cause of this is explained by the condensation behavior of $H_2O$-ice from gas phase into solids. It is a sharp feature: if conditions are right, within just a few degree Kelvin, all $H_2O$-ice will condense out, so one either gets most of it or very few of it (J. ~S. Lewis, 1997; J. S. Lewis & S., 1972; Lodders & Fegley, 2010; Zeng, Jacobsen, Sasselov, Vanderburg, et al., 2018). In fact, it must be because the earlier condensed dust will be entrapped into more abundant ices unless dust-gas fractionation has occurred before condensation of ices (Boogert, Gerakines, & Whittet, 2015; Oberg et al., 2011). If a planet core forms beyond or near snowline, it will grow from accreting rocks and ices simultaneously. The core can subsequently migrate in disk to become closer to host star (Kley & Nelson, 2012). Alternatively, a growing planet core can receive ices from pebbles which cross into the snowline (Johansen & Lambrechts, 2017).

For first scenario, both the rocky cores (formed initially inside the $H_2O$-snowline) and the icy cores (formed initially near or outside the $H_2O$-snowline) experience significant inward migration. They are revealed to us through the transit surveys such as *Kepler* and *TESS* since they are biased towards finding closer-in planets. If both types of cores exist by formation, then it is hard to imagine that migration only works for one but not the other (Raymond, Boulet, Izidoro, Esteves, & Bitsch, 2018).

Radial-velocity surveys suggest that the frequency of a system with inner low mass planets in the presence of outer giant planet is low ($\lesssim 10\%$). This is used as evidence supporting migration to explain both the super-Earth (1-2 $R_\oplus$) population and the mini-Neptune (2-4 $R_\oplus$) population (Barbato et al., 2018). This low frequency of solar-system analogs with long-period outer giant planet can be reproduced by population synthesis approach (Chambers & E., 2017). It has been speculated that there is a peak in the inward migration efficiency for the intermediate core masses which are in between the Type I and Type II regimes (Armitage, 2010). If so, this maximum migration efficiency could lead to the fast inward-migration of many rocky cores and icy cores of the order of ~10 $M_\oplus$ before they can accrete enough gas to grow into gas giants. Therefore, we speculate that around solar-type FGK stars:
In ~90% of planetary systems, migration overtakes growth, so that multiple cores migrate inward fast enough to form these super-Earths and mini-Neptunes that we observe, before they can accrete enough gas.

In ~10% of planetary systems, growth overtakes migration, so one core quickly grows into Jupiter-mass and opens up a big gap is the disk and dominates the system. It then stops any other cores from migrating inward through the disk.

## Summary

In summary, the survival function analysis provides a model-independent way to assess and classify different regimes of planets according to their sizes. The boundaries identified at 4 **$R_\oplus$** and 10 **$R_\oplus$** provide constraints in addition to the planet radius gap at 2 **$R_\oplus$** that any model of planet formation or interior structure should satisfy. We confirm that the slope of survival function in between 4 and 10 $R_\oplus$ to be shallower compared to either end, indicating a relative paucity of planets in between 4-10 $R_\oplus$, namely, the sub-Saturnian desert there. We name them transitional planets, as they form a bridge between the gas-poor small planets and gas-rich giant planets. Furthermore, we argue that planets in between 2-4 $R_\oplus$ are most likely water-rich cores, namely, water worlds, based on the comparison of their distribution in flux-dimension with the planet population just below the radius gap, and based on the comparison of their distribution with the prediction of gas-dwarf-evaporation-by-host-star hypothesis, based on the observations of the dichotomy of rocky versus icy bodies in our own solar system, and also based on the competition of growth-versus-migration of planet cores during planet formation in disk. This new interpretation of the planet radius gap and planets of 2-4 $R_\oplus$ as water worlds shall promote further observational test, including the atmosphere characterization of some of them through *JWST*, and more mass-radius measurements from *TESS* and its follow-ups.

**Acknowledgements**: This work was partly supported by a grant from the Simons Foundation (SCOL [award #337090] to L.Z.). Part of this research was also conducted under the Sandia Z Fundamental Science Program and supported by the Department of Energy National Nuclear Security Administration under Award Numbers DE-NA0001804 and DE-NA0002937 to S. B. Jacobsen (PI) with Harvard University. This research is the authors' views and not those of the DOE. Sandia National Laboratories is a multimission laboratory managed and operated by National Technology and Engineering Solutions of Sandia, LLC., a wholly owned subsidiary of Honeywell International, Inc., for the U.S. Department of Energy's National Nuclear Security Administration under contract DE-NA-0003525. We want to acknowledge Travis A. Berger, Daniel Huber, Eric Gaidos, and Jennifer L. van Saders for sharing their data in the *Revised Radii of Kepler Stars and Planets using GAIA Data Release 2* to aid our analysis. The author L.Z. would also like to thank Mercedes Lopaz-Morales, Juan Perez-Mercader, Michail I. Petaev, Thomas R. Mattsson, Raphaëlle D. Haywood, David W. Latham, Samuel Hadden, Matthew Heising, Gongjie Li, Jingjing Chen, Hao Cao, Eric J. Klobas, Mario Damasso, Aldo S. Bonomo,

Laura R. Kreidberg, Hilke E. Schlitchting, Lisa Kaltenegger, Morris Podolak, Amit Levi, Jordan Steckloff, and Eugenia Hyung for inspirational discussions.

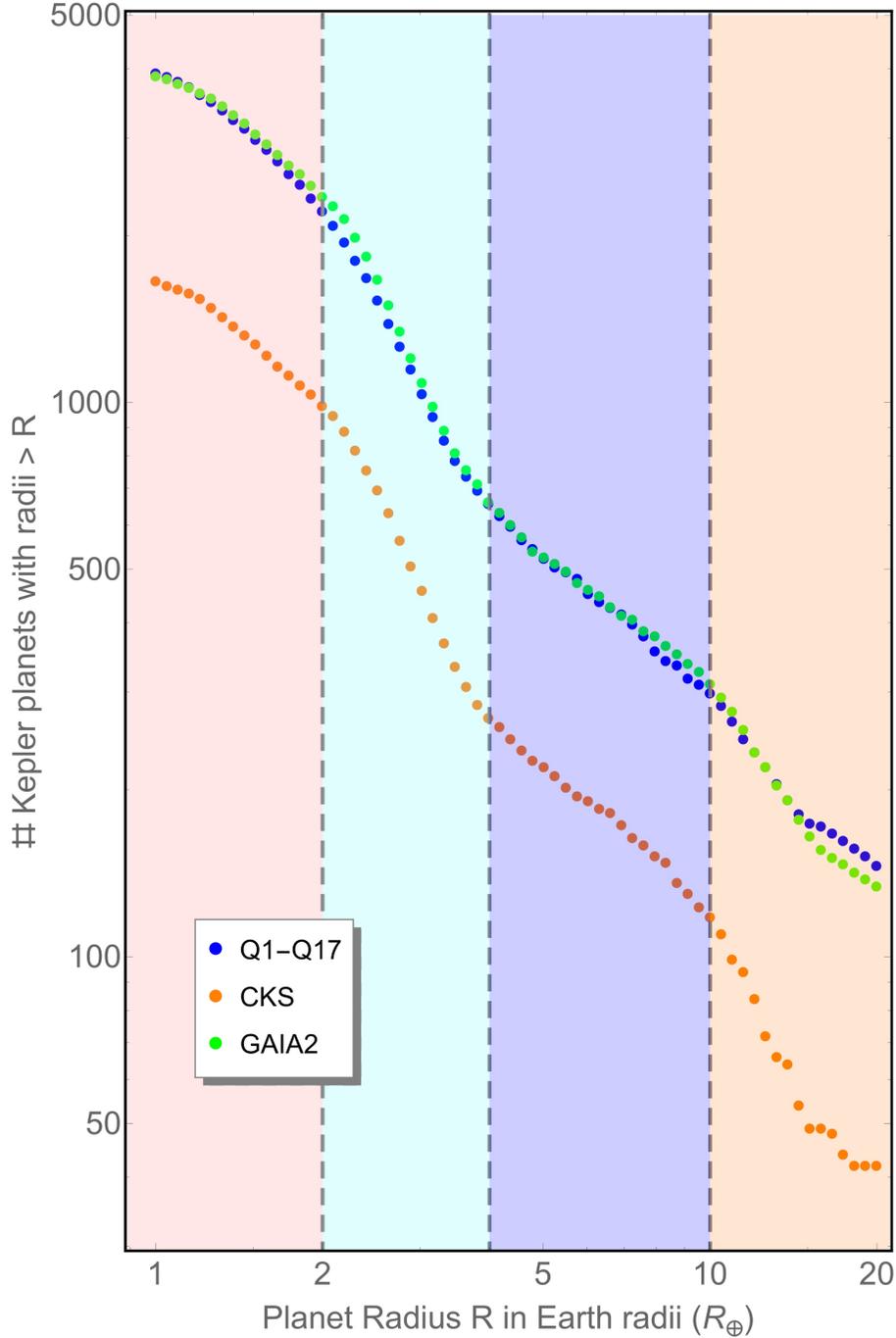

**Figure 1.** Survival function of planet radius of the *Kepler* confirmed/candidate planets (4433 from Q1-Q17 Data Release 25 of *NASA* Exoplanet Archive (Akeson et al., 2013; Thompson et al., 2017), 1861 from California-Kepler Survey (Fulton et al., 2017) with improved stellar parameters, both with false positives excluded already), and 4268 from the improved planet radii from GAIA Data Release 2 (Data courtesy of Travis Berger and Daniel Huber (Berger et al., 2018; Gaia Collaboration et al., 2018; Lindegren et al., 2018), where the typical errorbar of each planet radius is reduced to 5~10%. The proposed boundaries of different planet regimes are shown as vertical dashed lines.

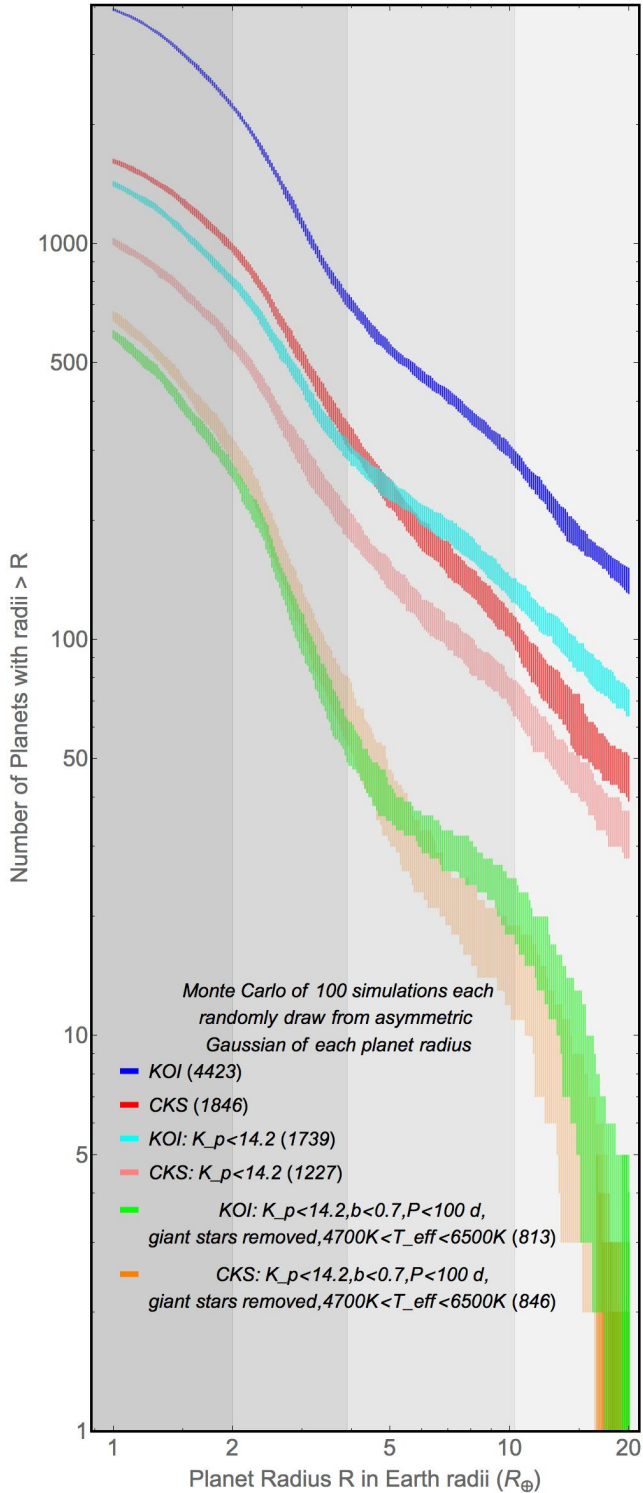

**Figure 2**. Survival function after applying successive cuts in conformity with the cuts in (Fulton et al., 2017) and taking into account error in planet radius using the Monte-Carlo method by drawing randomly from the Gaussian with asymmetric uncertainty on each side, centered at the best-fit value of radius of each planet, and plot the survival function. We repeat the whole process 100 times, and calculate 100 survival functions for each scenario, to give an idea as to the uncertainty in the survival function itself.

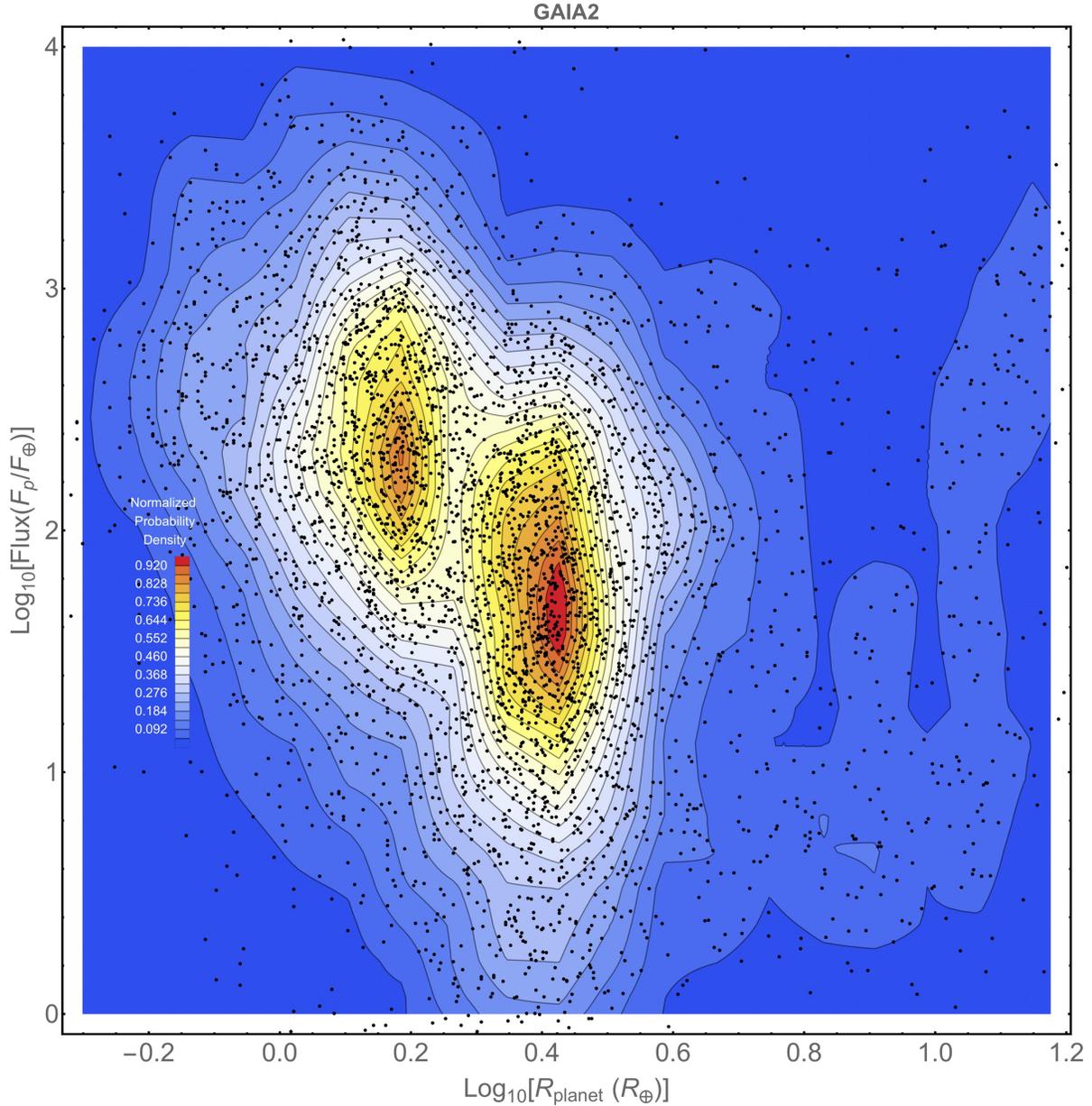

**Figure 3**. Smooth Kernel 2D-Histogram produced with *Mathematica*, showing the overall distribution of the *Kepler* confirmed/candidate planets in the $Log_{10}$[Planet Radius ($R_\oplus$)]-$Log_{10}$[Flux ($F_\oplus$)]-space. Data courtesy of ***Travis Berger*** and ***Daniel Huber***, from the Institute for Astronomy, University of Hawaii. The typical errorbar in the radius of each individual planet is about 5~10% here (Berger et al., 2018). The radius gap at $2=10^{0.3}$ Earth radii ($R_\oplus$) is perpendicular to the radius-axis. Notice that the gap is partially filled-in (Fulton & Petigura, 2018), contradictory to earlier speculations of an observationally under-resolved and forbidden region of planet sizes (Fulton et al., 2017; Van Eylen et al., 2017). Also, it looks as if the gap extends down to but disappears around ~100 Earth fluxes ($F_\oplus$). This is likely an artifact due to geometric transit probability. It is not because the gap does not exist at lower flux or larger orbital semi-major axis or longer orbital period, but because the geometric transit probability decreases as (1/**a**) where **a** is the orbit semi-major axis, so it looks as if the population below the gap diminishes and the gap vanishes. This point will be elaborated in forthcoming analysis that once the geometric transit probability is corrected, both planet populations (1-2 $R_\oplus$ and 2-4 $R_\oplus$) are nearly flat (log-uniform) from the inner cut-off (the attenuation of both populations towards higher fluxes) out to at least 1 AU or 1 $F_\oplus$. The break at $4=10^{0.6}$ $R_\oplus$ can be viewed as the rapid falloff of the population of 2-4 $R_\oplus$ towards larger radius. The break at 10 $R_\oplus$ can be viewed as the upper bound of the sub-Saturnian desert delineated nicely by one of the contour lines.

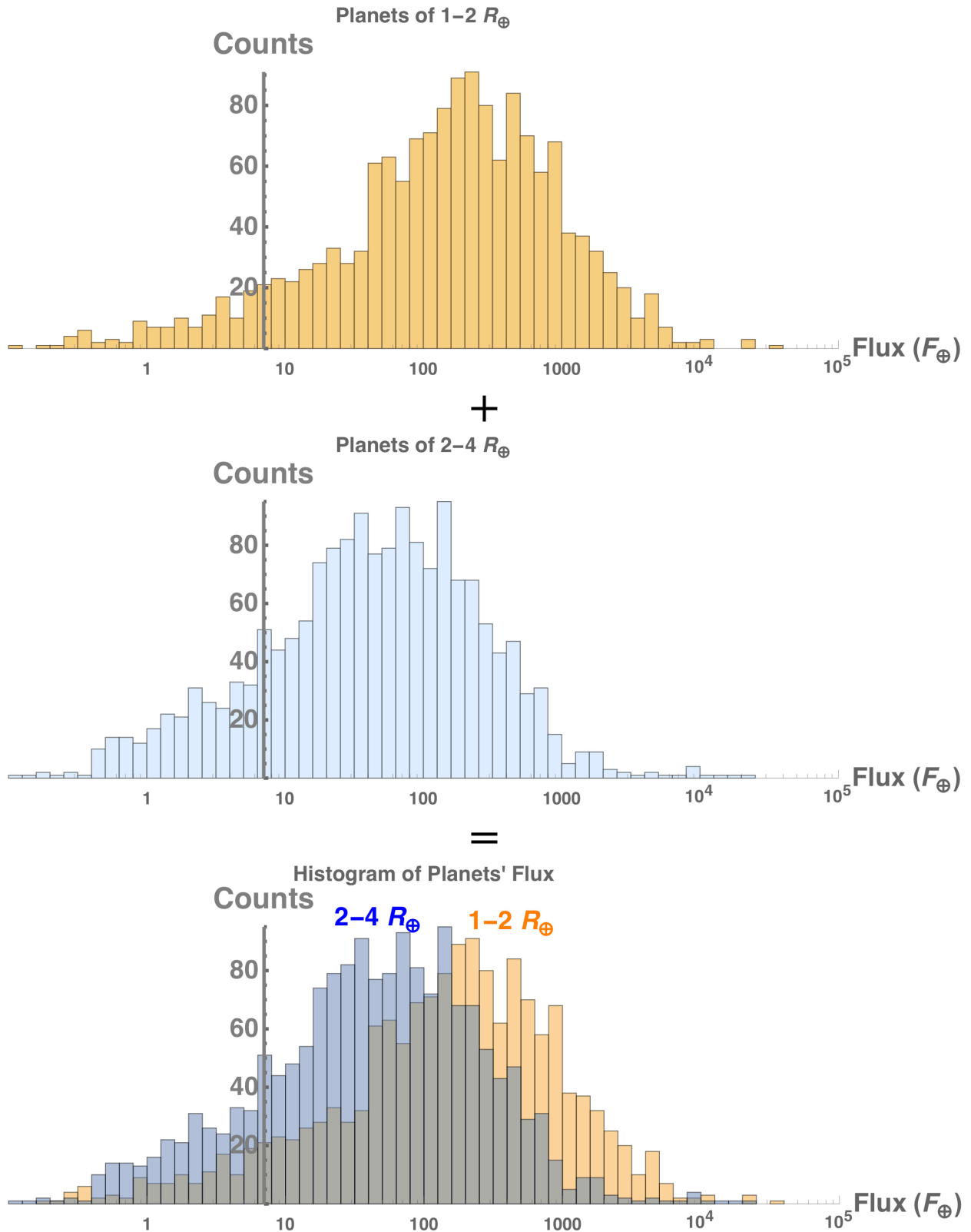

**Figure 4**. The histograms of flux show the attenuation of both planet populations with increasing flux. Interestingly enough, the two planet populations (1-2 $R_\oplus$ vs. 2-4 $R_\oplus$) overlap significantly in the flux-dimension and show similar trends of attenuation towards higher flux, except shifted in flux by a factor of 3~4.

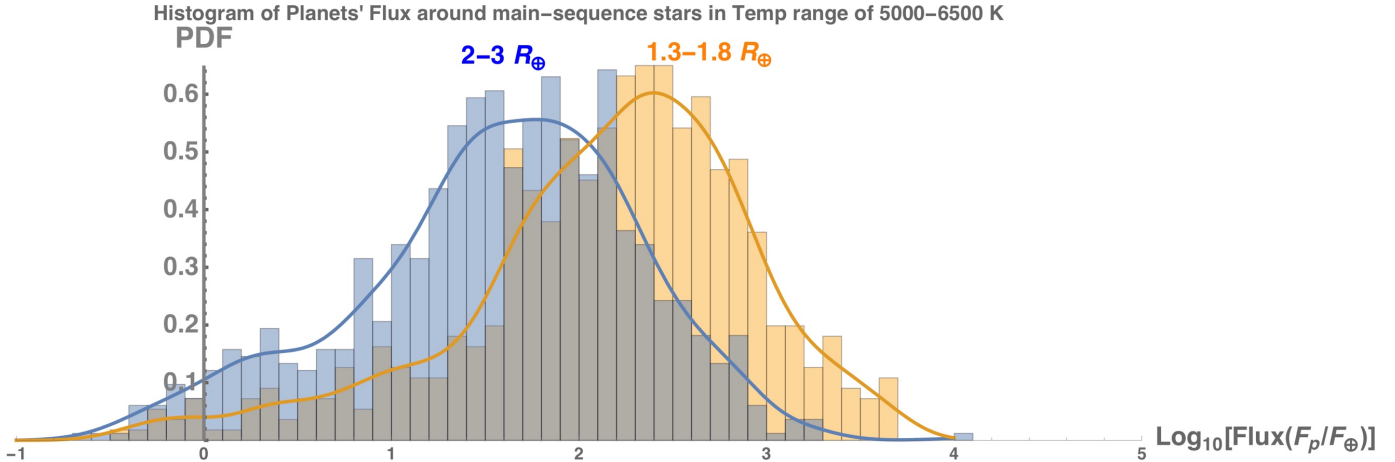

**Figure 5**. The probability density function (PDF) in flux of the two populations (1.3-1.8 R⊕ vs. 2-3 R⊕) of the *Kepler* planets/candidates from the GAIA DR2 updates (Berger et al., 2018). Each PDF is fitted to a smooth curve using *Mathematica*'s built-in function: "SmoothKernelDistribution" using the "Sheather-Jones" bandwidth selection method. It is worth mentioning that, the planets of 1.3-1.8 R⊕ have abrupt and complete cut-off at 5000 F⊕, and except for one outlier, the planets of 2-3 R⊕ have a complete cut-off at 2000 F⊕.

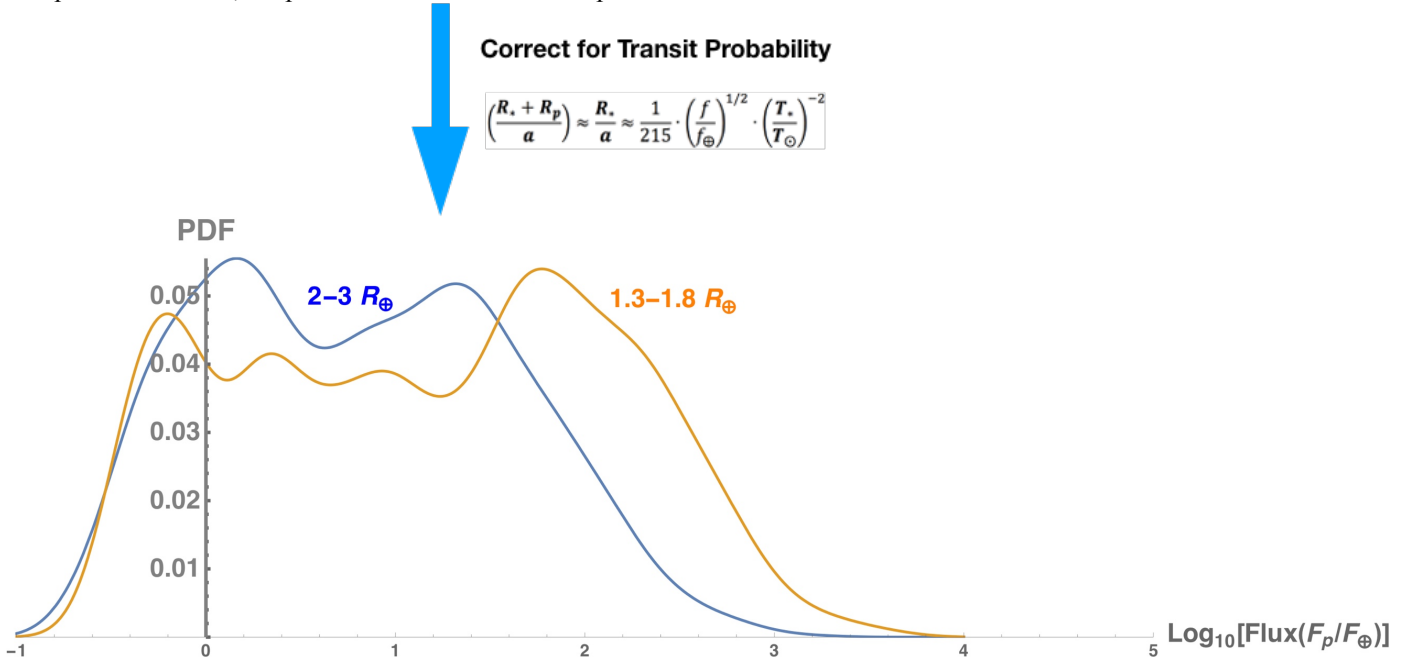

**Figure 6**. The transit-probability-corrected PDF in log-flux-dimension of the two planet populations (1.3-1.8 R⊕ vs. 2-3 R⊕) of the *Kepler* confirmed/candidate planets with GAIA DR2 updates (Berger et al., 2018). Notice that the geometric shape of the PDF attenuation towards higher flux is very similar between the two populations, except shifted in flux by a factor of 3~4. Moreover, the un-attenuated part of the PDF in the low-flux region is nearly flat (log-uniform) in flux (and equivalently semi-major axis and orbital period), confirming the results revealed by the survival function analysis of planer orbit semi-major axis distribution (Zeng, Jacobsen, Sasselov, & Vanderburg, 2018). On the other hand, the attenuation of PDF to the left of ~1 F⊕ is primarily due to the limited observational duration (~3.5 years) of the *Kepler* mission before *K2*. The amplitude of the blue curve is manually reduced by a factor of 2 in order to match the height of the orange curve to be compared with. This suggests that the intrinsic PDF (dN/dlog(flux)) of planets of (2-3 R⊕) is roughly twice that of the PDF of planets of (1.3-1.8 R⊕) in the flat un-attenuated regime. Furthermore, the pipeline incompleteness correction, if applied, will increase the number of small planets of (1.3-1.8 R⊕) in the low-flux (≤100 F⊕) region even slightly more. The gas-dwarf-evaporation-by-host-star model (see **Figure 7**) has difficulty explaining the persistence of this population of planets in between (1.3-1.8 R⊕) at low-flux/long-orbital-period.

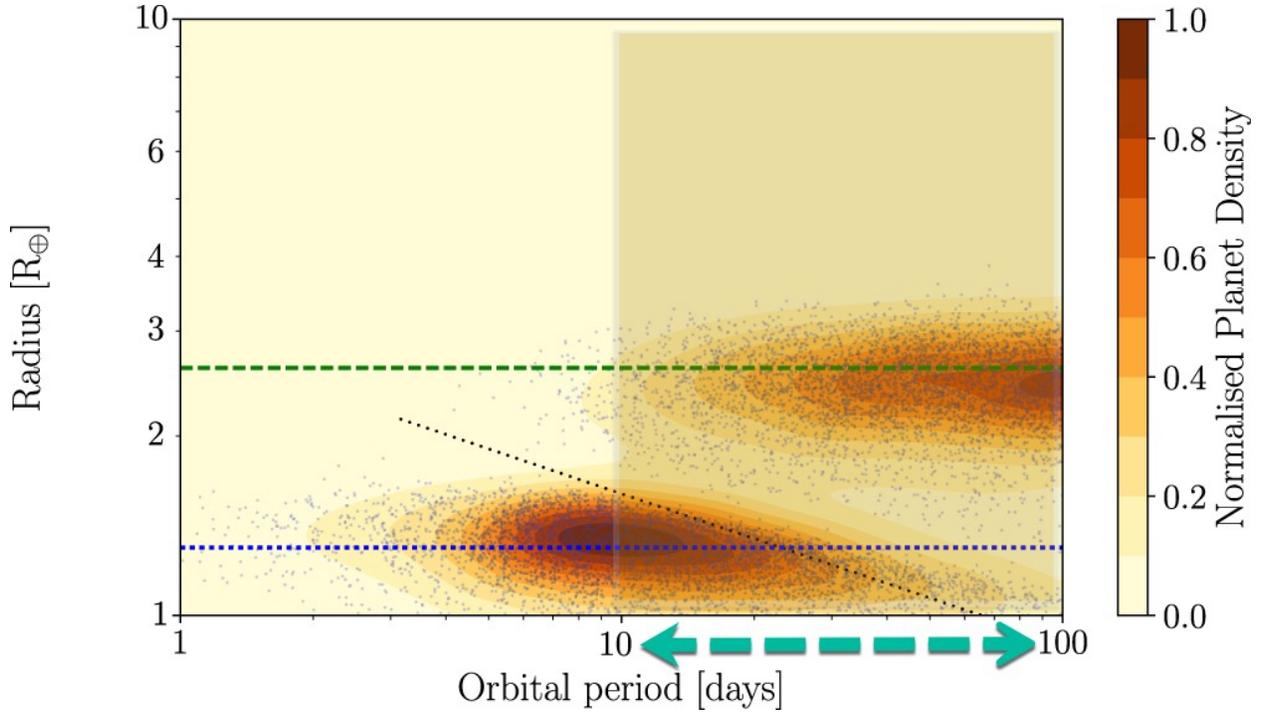

**Figure 7**. Modified from *Figure 9* of (Owen & Wu, 2017), where the authors have stretched their model to generate as many long-period small (rocky) planets as possible. Notice that the two planet populations below and above the radius gap at ~2 R$_\oplus$ created by photo-evaporation are complementary to one another, that is, the high in PDF of one is the low in PDF of the other, since they are assumed to be derived from the same intrinsic gas-dwarf population. Therefore, The ratio of PDFs between the two planet populations, below and above the gap, say PDF[(1.3-1.8 R$_\oplus$)/PDF[(2-3 R$_\oplus$)], becomes strongly dependent on orbital period, or equivalently flux or semi-major axis.

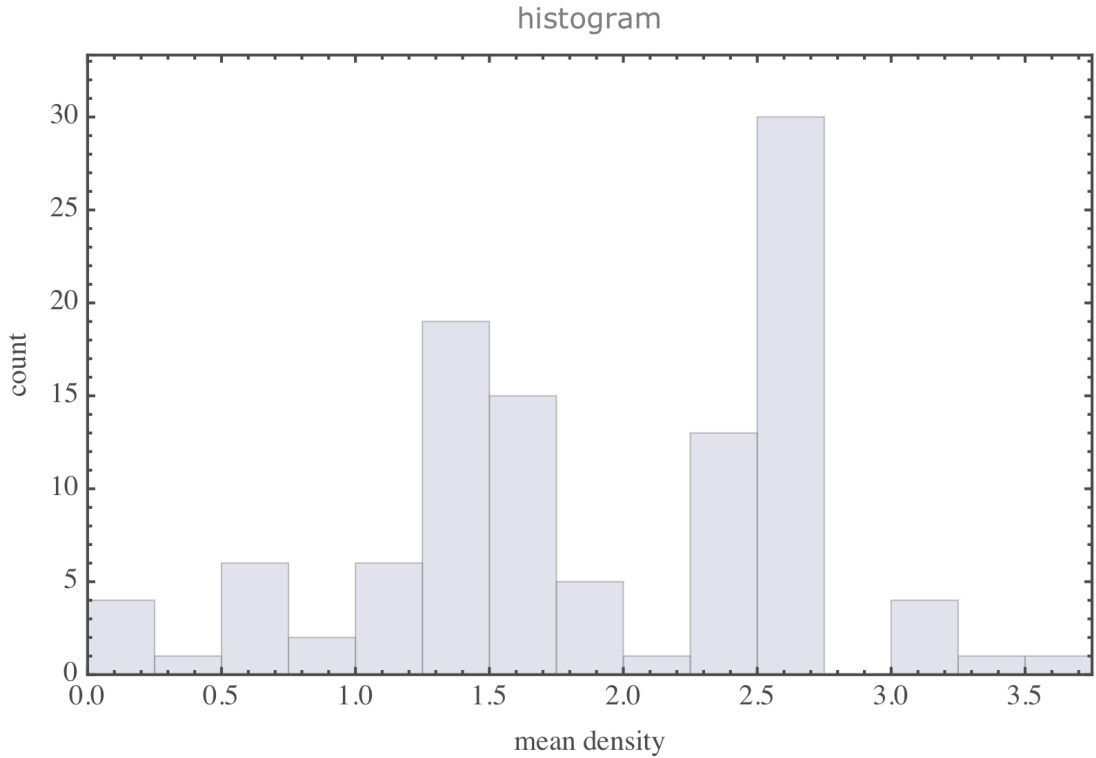

**Figure 8**. Histogram of densities (g/cc) of solar system satellites, from *WolframAlpha* ([www.wolframalpha.com](http://www.wolframalpha.com))